\documentclass{article}
\usepackage{spconf,amsmath,graphicx}
\usepackage{subfigure}
\usepackage{courier}  
\usepackage{url}            
\usepackage{amsfonts}       
\usepackage{nicefrac}       
\usepackage{xcolor}         
\usepackage{mathrsfs}
\usepackage{microtype}
\usepackage{multirow}
\usepackage{booktabs}
\usepackage{amsmath, amssymb}
\usepackage{mathrsfs}
\usepackage{subfigure}
\usepackage{pgf-pie}
\usepackage{pgfplots}
\pgfplotsset{compat=1.18}
\usepackage{caption} 
\usepackage{algorithm}
\usepackage{algorithmic}

\usepackage{newfloat}
\usepackage{listings}
\usepackage{multirow,multicol}
\usepackage{algorithm}
\usepackage{algorithmic}
\usepackage{algorithmic}\makeatletter

\title{Learning Disentangled Speech Representations with Contrastive Learning and Time-Invariant Retrieval}
\name{Yimin Deng$^{1,2\ddagger}$, Huaizhen Tang$^{1,3\ddagger}$\thanks{$^\ddagger$ Both authors have equal contributions.}, Xulong Zhang$^{1\ast}$, Ning Cheng$^{1\ast}$\thanks{$^\ast$Corresponding authors: Xulong Zhang (zhangxulong@ieee.org), Ning Cheng (chengning211@pingan.com.cn).}, Jing Xiao$^{1}$, Jianzong Wang$^{1}$ }
\address{$^{1}$Ping An Technology (Shenzhen) Co., Ltd., China\\$^{2}$University of Science and Technology of China\\$^{3}$Huya Inc (Shenzhen) Co., Ltd.}
\begin{document}
%
\maketitle
\begin{abstract}

Voice conversion refers to transferring speaker identity with well-preserved content. Better disentanglement of speech representations leads to better voice conversion. Recent studies have found that phonetic information from input audio has the potential ability to well represent content. Besides, the speaker-style modeling with pre-trained models making the process more complex. To tackle these issues, we introduce a new method named ``CTVC" which utilizes disentangled speech representations with contrastive learning and time-invariant retrieval. Specifically, a similarity-based compression module is used to facilitate a more intimate connection between the frame-level hidden features and linguistic information at phoneme-level. Additionally, a time-invariant retrieval is proposed for timbre extraction based on multiple segmentations and mutual information. Experimental results demonstrate that ``CTVC" outperforms previous studies and improves the sound quality and similarity of the converted results.

\end{abstract}
\section{Introduction}
Voice Conversion~(VC), also called ``voice style transfer'', the primary objective of which is to modify one's voice to resemble that of another preserving the linguistic content~\cite{study}. 
As an essential aspect of speech synthesis, wide-ranging applications of VC hold considerable significance in human-computer interaction, such as customer service, movie dubbing, and communication aids for those with speech impairments.

To achieve a satisfactory quality of conversion, voice conversion needs to disentangle and manipulate speaker-specific attributes such as timbre, emotion, and accents while maintaining the integrity of the linguistic information~\cite{chan2022speechsplit2,deng2023PMVC}. That means robust disentanglement of speech representations is necessary during this process. Then a decoder is trained to generate a natural speech from extracted speech representations. With a well-trained network, the timbre can be controlled by target speech and the content from source speech can be completely expressed in inference time.
Recent studies have been conducted on disentangled speech representations learning and made considerable progress. With a strong focus on maintaining linguistic content integrity, self-supervised learning methods~\cite{hubert,chen2022wavlm} have aroused public attention in this area~\cite{van2022softvc,li2023freevc}. Latent speaker information in the content representation may not be entirely eliminated, resulting in the failure of voice conversion. A common solution for the identity exchange is to embed a related vector from pre-trained voice print recognition models~\cite{d-vector,x-vector}.

However, introducing pre-trained models will increase the complexity and make it difficult to generalize to applications. AutoVC~\cite{autovc} proposes a basic framework with autoencoders. It encourages encoders to learn disentangled speech representations simultaneously and works well in an unseen corpus which is normally employed in real life. Derived from such framework, vector quantization~\cite{avqvc}, text encoder guidance~\cite{tgavc}, phonetic posteriorgram~\cite{PPG-VC2} and bottleneck features~\cite{zhao2022bnfs} are introduced to achieve better conversion. Inspired by these, we should pursue a more concise and elegant implementation to perform speech conversion tasks. 

\begin{figure*}[th]
    \vspace{-2em}
    \centering 

    \includegraphics[scale = 0.53]{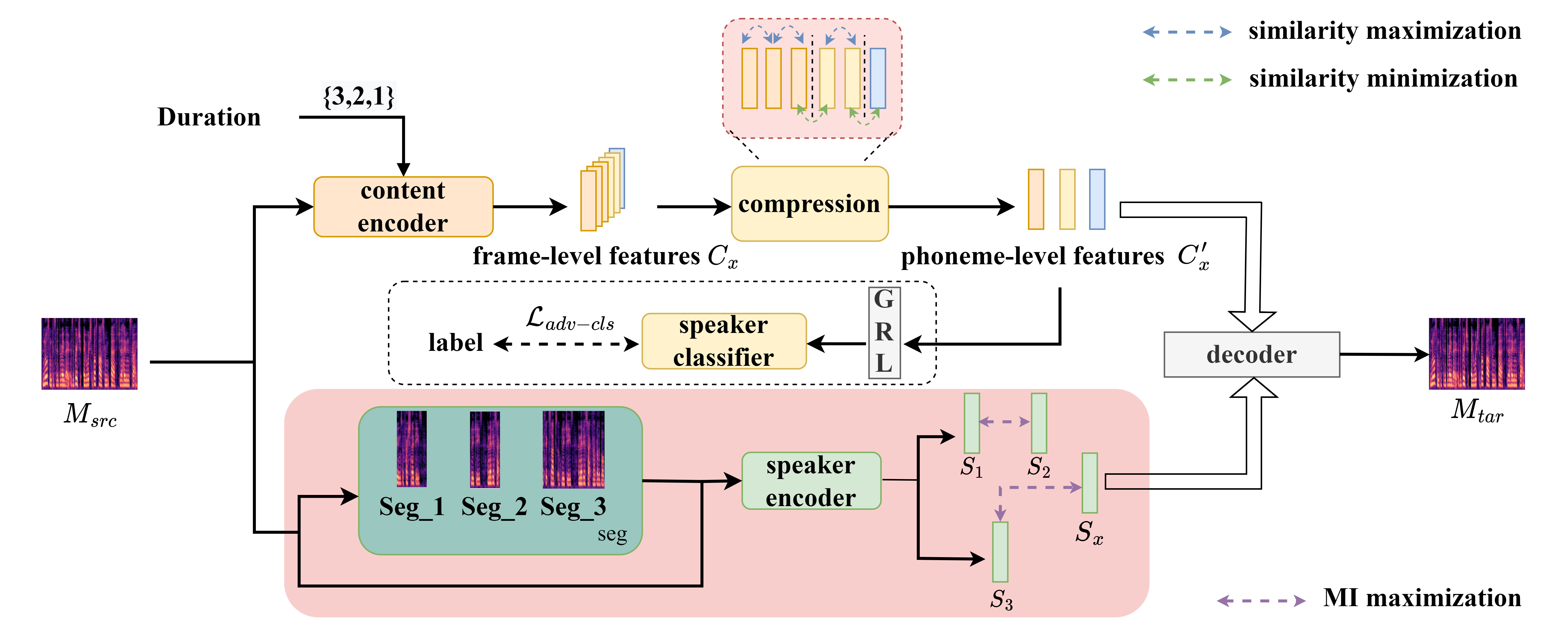}
    \vspace{-1em}
    \caption{The framework of ``CTVC". $C_x$ is the content embedding that is generated by the content encoder while $S_x$ refers to the global speaker embedding. \textbf{GRL} denotes Gradient Reversal Layer. \textbf{MI} means Mutual Information. In compression module, the colors indicates frames belonging to different phonemes and the dash-lines indicates boundaries.}
    \vspace{-1.5em}
    \label{fig:model-arch}
    
\end{figure*}

We propose a novel VC framework named ``CTVC" based on disentangled speech representation. Inspired by recent work~\cite{tian2020makes}, take advantage of some forced aligner tools like MFA~\cite{MFA} and we can get the duration sequence. Then, a similarity-based compression is designed to construct the ideal content features at the phoneme-level from the frame-level hidden speech representations. Considering the inter-frame similarity in the whole utterance, a novel approach based on time-invariant retrieval is also proposed for speaker representation learning.
The main contributions are as follows:

\begin{itemize}
    \item A novel training approach with contrastive similarity loss is employed to steer the content embedding towards purer linguistic information, while simultaneously excluding style information from the encoder output.
    \item A time-invariant retrieval method is designed to encourage the speaker representation to contain the global style while discarding the time-variant features.
\end{itemize}

\section{Methodology}
\subsection{Speaker-Independent Content Feature Discovery}
Initially, given an audio waveform and its corresponding T-frame mel-spectrogram $X=(x_1, x_2, ..., x_T)$, the content encoder $E_c$ acquires its hidden feature at frame-level $C_X = E_c(X) = (c_1, c_2, ..., c_T)$. Then, with the duration sequence from forced alignment, for each pair of frame indexes $(i, j)$, it's straightforward to know whether $c_{i}$ and $c_j$ share identical content information or not as shown in Fig.~\ref{fig:model-arch}. Then use a score as measurement for feature similarity between a pair of frames in the form of cosine similarity:

\begin{align}
    G(c(x_{i}), c(x_{j})) = \frac{c^T(x_{i})c(x_{j})}{\|c(x_{i})\|_2\|c(x_{j})\|_2}
    \vspace{-0.5em}
\end{align}
where $c(\cdot)$ denotes the extracted hidden representation of specific frame. It's expected that the similarity is high between hidden representations of the same phoneme while it's low between the hidden features at the boundary as shown in compression module in Fig.~\ref{fig:model-arch}.
Therefore, the similarity contrastive loss function is:
\begin{equation}
    \mathcal{L}_{\text{sim}} = \sum_m^{M} \sum_{i,j=1}^{T} (-1)^{h} G(E_c(x_i), E_c(x_{j})), 
    \vspace{-0.5em}
\end{equation}
where $i, j$ denotes the $i^{th}$ and $j^{th}$ frame of mel-spectrogram. $h$ equals 1 when both frames belong to the same phoneme while -1 for those from different ones. $T$ denotes the number of frames, $M$ indicates the quantity of contrastive samples. During training, the minimizing of contrastive loss $\mathcal{L_\text{sim}}$ will force the content encoder to generate the frame-level content embedding that are closely associated with the linguistic information.


To further elminate speaker information, we apply domain adversarial training. A Gradient Reversal Layer (GRL) is positioned between content encoder and speaker domain classifier. During training, the content features $C_{x}$ is fed into an auxiliary speaker domain classifier to predict the speaker identity. The speaker classifier would be expected to perform as accurately as possible. To eliminate speaker information, due to the GRL, the content encoder has the opposite optimization goal with the adversarial loss: 
\begin{align}
    \label{eq:4}
    \mathcal{L}_{\text {adv-cls }}(\boldsymbol{\theta_e, \theta_{cls}}) &= -\sum_{k=1}^K\mathbb{I}(Spk_u==k) \log p_k
\end{align}
where $\mathbb{I}(\cdot)$ functions as an indicator that whether the speaker $Spk_u$ producing speech $u$ is speaker $k$. As there are $K$ speakers totally. Use $p_k$ to represent the probability corresponding to specific speaker.
$\boldsymbol{\theta_{cls}}$ and $\boldsymbol{\theta_{e}}$ are learnable parameters of speaker classifier and content encoder respectively. During training, with $\mathcal{L}_{\text{adv-cls}}$, $\boldsymbol{\theta_{cls}}$ are optimized to better identify the corresponding speaker. Simultaneously, $\boldsymbol{\theta_{e}}$ are optimized to deceive the speaker classifier. Ideally, under the constraint, the content encoder's output will discard speaker style and cannot be used to distinguish speakers.

\subsection{Time-Invariant Retrieval for Speaker Representation}

        
In this section, we will focus on the representation of timbre information. The speaker encoders in previous work are often with a structure of CNN and pooling layers, which intends to learn global information of the speech. However, it lacks effective constraint to get rid of latent time-relevant information that may have an impact on phonemes.

As we assumed above, the global style representation is independent of the time axis, which comes from our life experience: for each utterance, we don't need to listen to the whole speech but only need to listen to a part of the speech to judge the identity of the speaker. Based on this assumption, we design a time-invariant retrieval to encourage our model to learn global style embedding closer to the ideal one.

\begin{figure}[htp]
    \vspace{-1.5em}
    
    \subfigure[Cut two segments]{
        \label{style_1}
        \centering          
		\includegraphics[scale=0.2]{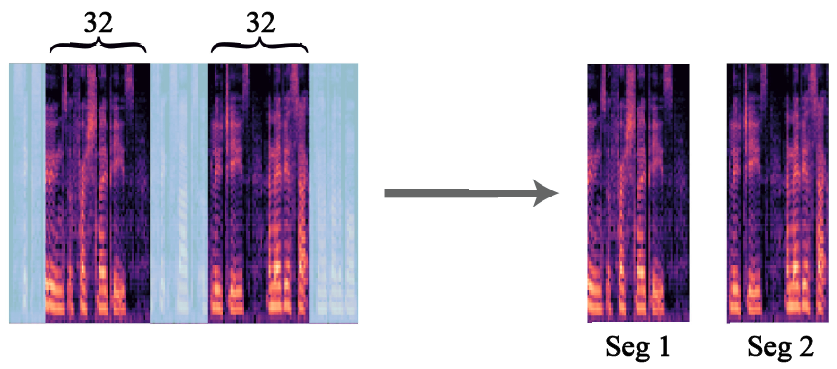}  
			
  
    }
    \hfill
    \subfigure[Entire and part]{
        \label{style_2}
        \centering      
		\includegraphics[scale=0.22]{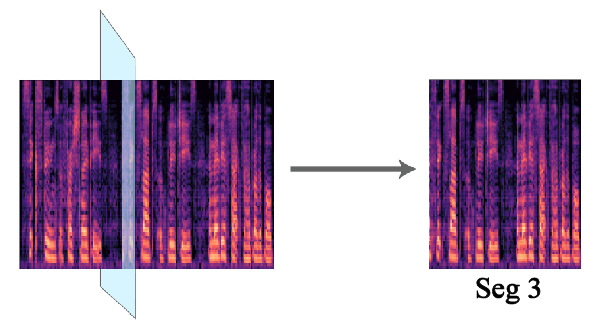}  
			
        
    }
    \caption{Different segment methods for style controlling}
    \label{fig:style-control}
\end{figure}

As illustrated in Fig.~\ref{fig:style-control}, we expand the description of the seg part in Fig.~\ref{fig:model-arch}. We first try to randomly intercept a 32-frame speech segment from the first half and the second half of the speech respectively, then we will get two 32-frame speech segments $Seg1$ and $Seg2$ shown in Fig.~\ref{fig:style-control}. According to our hypothesis, the content information is time-variant, while the speaker information is time-invariant. Hence, we expect the style embeddings $s_1$ and $s_2$ should be as similar as possible. 
The correlation between the two segments needs to be measured. In order to more comprehensively capture the correlation between features, especially in complex data relationships, we use Mutual Information~(MI) here. Given the random variables $u$ and $v$, the MI is Kullback-Leibler (KL) divergence between their joint and marginal distributions as:
 
 \begin{equation}
     I(u, v) = D_{KL} (P(u, v); P(u)P(v))
 \end{equation}
Following the previous work~\cite{MI2021}, MI maximization tasks correspond to the maximization of the lower bound of MI. We employ a widely-used lower bound InfoNCE~\cite{infonce}:
\begin{equation}
\mathcal{I}_{\mathrm{NCE}}(\boldsymbol{u},\boldsymbol{v})=\mathbb{E}\left[\frac{1}{N} \sum_{i=1}^N \log \frac{e^{f\left(\boldsymbol{u}_i, \boldsymbol{v}_i\right)}}{\frac{1}{N} \sum_{j=1}^N e^{f\left(\boldsymbol{u}_j, \boldsymbol{v}_j\right)}}\right]
\end{equation}
where $\boldsymbol{u},\boldsymbol{v}$ denote the $s_1$ and $s_2$ style embeddings of the segments, with a score function $f(u, v)$ based on a simple log-bilinear model~\cite{infonce}.

\begin{equation}
    f(\boldsymbol{u}_i,\boldsymbol{v}_i) = \exp(\boldsymbol{h}^t_iW_i\boldsymbol{v}_i)
\end{equation}
For speaker $i$, $\boldsymbol{h}^t_i$ is the latent representation from input $\boldsymbol{u_i}$ and $W_i$ is a linear transformation used with $\boldsymbol{v_i}$ for prediction. 
Besides, we've tried another way to obtain the ideal style embedding. As shown in Fig.~\ref{style_2}, we randomly intercept more than half of the speech segment $Seg3$ from the whole speech $x$. Based on the same assumption, we think their style embeddings are still very similar. In the training phase, the style loss can be updated as:

\begin{align}
    \mathcal{L}_{s} =& \sum_i^{N} I(s_{1,i},sg(s_{2,i}))+\sum_i^{N} I(s_{2,i},sg(s_{1,i})) \nonumber \\
                     & + \sum_i^{N} I(s_{x,i},sg(s_{3,i}))+\sum_i^{N} I(s_{3,i},sg(s_{x,i}))
\end{align}

\noindent where $s_1$, $s_2$, $s_3$ and $s_x$ denote the style embedding of the corresponding speech segments from speaker $i$. $sg$ indicates the stop-gradient operation, $N$ is the number of speakers.  With the Time-Invariant Retrieval strategy, the speaker encoder $E_s$ is forced to retrieve the time-invariant global style information. 
\subsection{Training Strategy}
The concatenation of the content encoder outputs and style encoder outputs are input into the decoder module and output the predicted mel spectrogram, where the reconstruction loss is calculated between the target mel $x_n$ and predicted mel $\hat{x_n}$ of utterance $u$ where $N_u$ is the number of utterances.
\begin{equation}
    \mathcal{L}_{recon} = \sum_n^{N_u} ||(x_n, \hat x_n)||^2_2
\end{equation}
The loss functions involved in training are as follows: 

\begin{align}
    \label{full loss}
    \mathcal{L}(\boldsymbol{\theta_{e_c}, \theta_{e_s}, \theta_d,\theta_{cls}}) = \mathcal{L}_{\text {recon}} + \alpha \mathcal{L}_{\text {sim}} + \beta \mathcal{L}_{\text {s}} + \lambda \mathcal{L}_{\text {adv-cls}}
\end{align}
where the constant coefficients $\alpha$ , $\beta$, and $\lambda$ refer to the weights of different loss functions respectively. $\theta_{e_c}$, $\theta_{e_s}$, $\theta_d$ and $\theta_{cls}$ are regularization parameters of the content encoder, speaker encoder, decoder and classifier. With this objective loss function, well-constructed speech representations can be learned.

\section{Experiments}

\begin{table*}[htbp]

    \vspace{-1.5em}
  \caption{Subjective and objective evaluations results in Many-to-Many and One-Shot Voice Conversion tasks}
  \vspace{-0.5em}
  \centering
  \fontsize{8.7}{7}\selectfont
  \label{Comparison}
    \begin{tabular}{ccccccc}
    
    \toprule
    \multirow{2}{*}{\textbf{Methods}}&
    \multicolumn{3}{c}{\textbf{Many-to-Many VC}}&\multicolumn{3}{c}{\textbf{ One-Shot VC}}\cr
    \cmidrule(lr){2-4} \cmidrule(lr){5-7}
    & MCD$\downarrow$ & MOS$\uparrow$ & VSS$\uparrow$ & MCD$\downarrow$ & MOS$\uparrow$ & VSS$\uparrow$\cr
    \midrule
    VQVC+~\cite{vqvc+} & 7.48 $\pm$ 0.26 & 2.89 $\pm$ 0.33 
    & 3.21 $\pm$ 0.46 & 8.41 $\pm$ 0.08 
    & 3.39 $\pm$ 0.36 & 3.08 $\pm$ 0.25 \cr
    F0-AutoVC~\cite{f0-autovc}& 8.34 $\pm$ 0.12 & 3.42 $\pm$ 0.37
    & 3.19 $\pm$ 0.48 & 8.66 $\pm$ 0.17 
    & 3.42 $\pm$ 0.73 & 3.31 $\pm$ 0.46 \cr
    ClsVC~\cite{Tang2023ClsVC} & 7.63 $\pm$ 0.18
    & 3.67 $\pm$ 0.52 
    & 3.13 $\pm$ 0.36 & 8.22 $\pm$ 0.31
    & 3.51 $\pm$ 0.50 & 3.08 $\pm$ 0.44 \cr
    TGAVC~\cite{tgavc}& 8.28 $\pm$ 0.11 & 3.59 $\pm$ 0.51
    & 3.28 $\pm$ 0.49 & 8.33 $\pm$ 0.09 & 3.58 $\pm$ 0.28 & 3.24 $\pm$ 0.40 \cr
    \midrule
    TVC & 7.33 $\pm$ 0.22 & 3.73 $\pm$ 0.41 
    & 3.12 $\pm$ 0.47 & 7.64 $\pm$ 0.28 
    & 3.60 $\pm$ 0.27 & 3.02 $\pm$ 0.25 \cr
    \textbf{CTVC} &\textbf{7.31 $\pm$ 0.19} &\textbf{3.75 $\pm$ 0.44}
    &\textbf{3.71 $\pm$ 0.42} & \textbf{7.62 $\pm$ 0.15} 
    &\textbf{3.66 $\pm$ 0.54} &\textbf{3.42 $\pm$ 0.39} \cr
    \bottomrule
    \end{tabular}
    \vspace{-1.5em}
\end{table*}
\subsection{Datasets and Configurations}

We conduct the objective and subjective experiments on Many-to-Many and One-Shot VC tasks for evaluation of model performance. Use a multi-speaker corpus, AISHELL-3~\cite{shi2020aishell3}, which containing 88035 recordings (roughly 85 hours) from 218 native Mandarin speakers. 
Select the recordings from 180 speakers for training and testing. The unseen voices of other speakers are employed for One-Shot test.
In Eq.(\ref{full loss}), the weights  are hyperparameters: $ \alpha = 0.01, \beta = -0.1, \lambda = 0.5$. Select F0-AutoVC~\cite{f0-autovc}, ClsVC~\cite{Tang2023ClsVC}, TGAVC~\cite{tgavc}, and VQVC+~\cite{vqvc+} as baseline models. Besides, to test the compression module and phonme-level feature, we use discrete speech units from HuBert for content extraction and remove the compression with contrast loss and retrain the model named ``TVC". We use a pre-trained high fidelity vocoder~\cite{hifigan} to transfer the mel-spectrums into waveform for listening tests.
\subsection{Comparison of VC Tasks}
In objective tests, the Mel-Cepstral Distortion~(MCD) is used to measure the difference between converted spectral features and target ones. The lower MCD means better performance. In subjective tests, conduct a listening tests with Mean Opinion Score~(MOS) to evaluate the sound quality. 13 volunteers( 8 males and 5 females) are invited to rate a score from 1-5 points on the naturalness of the results. Besides, subjects also take a voice similarity score~(VSS) test to measure the similarity between the converted voice and ground truth. Both are higher for better. We evaluate the performance of ``CTVC'' in different VC tasks. As shown in Table 1, in Many-to-Many VC, our model performs better than baselines in spectrum conversion and human perception. It's also shown that our model still performs well even for the unseen speakers during training process which achieves a lower MCD value and higher scores in naturalness and similarity.

\subsection{Evaluation of Speaker Similarity}
\begin{figure}[!t]
    \centering
    \subfigure[F-F]{
        \label{same-gender conversion 1}
        \begin{minipage}[b]{0.47\linewidth}
            \includegraphics[width=1\textwidth]{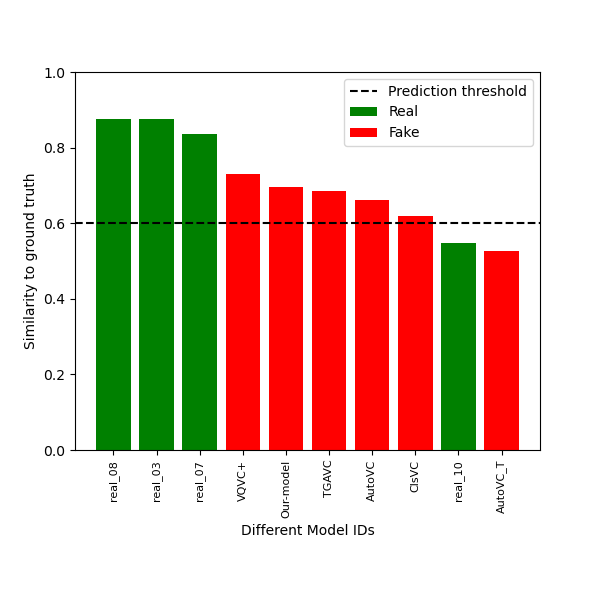}
        \end{minipage}
    } 
    \vspace{-1em}
    \subfigure[F-M]{
         \label{cross-gender conversion 1}
        \begin{minipage}[b]{0.47\linewidth}
            \includegraphics[width=1\textwidth]{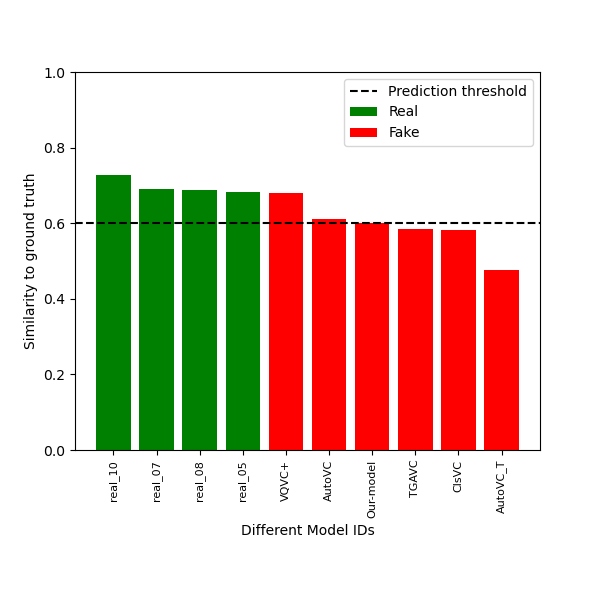}
        \end{minipage}
    }
    \vspace{-1em}
    \subfigure[M-M]{
        \label{same-gender conversion 2}
        \begin{minipage}[b]{0.47\linewidth}
            \includegraphics[width=1\textwidth]{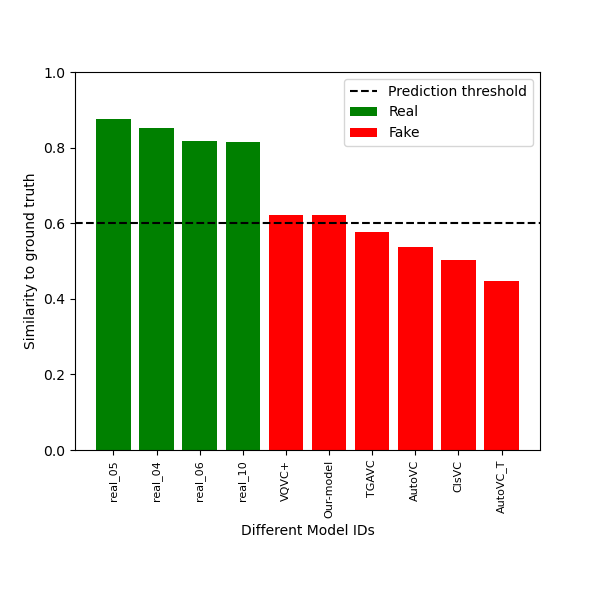}
        \end{minipage}
    }
    \subfigure[M-F]{
        \label{cross-gender conversion 2}
        \begin{minipage}[b]{0.47\linewidth}
            \includegraphics[width=1\textwidth]{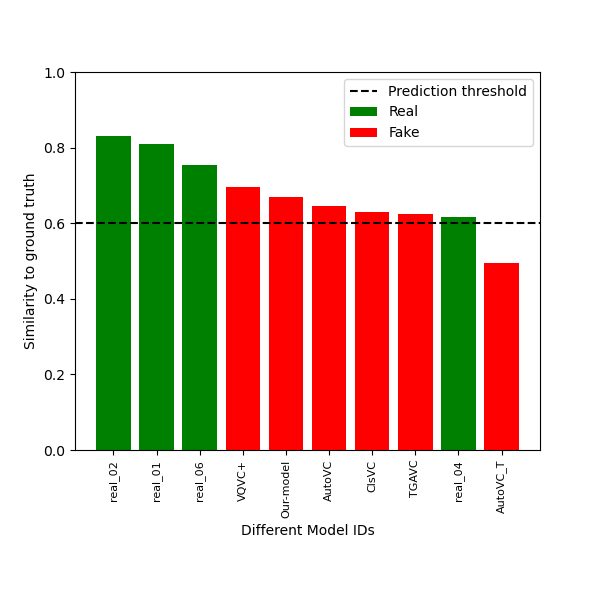}
        \end{minipage}
    }
    
    \caption{Objective evaluation results for Voice Conversion. F: Female; M: Male. Green groups are real speech. Red groups are synthesized speech from different models.}
    \label{fake}
\end{figure}

To conduct further objective evaluation of different converted models, we apply an open-source speech detection toolkit, \textit{Resemblyzer} to evalute the voice similarity.
Detaily, to compare the voice similarity between converted results and real audio, it will give a score ranging from 0 to 1. A higher score signifies a greater similar between the fake voice and the real voice. In addition, we repeat this experiment 20 times, so the final score will tend to be similar to the score of the target speaker's timbre. As shown in Fig.~\ref{fake}, the dash-line indicates the score that passes the toolkit test. Our model achieve a better stage in fake detection and outperforms than baselines in voice similarity.


\subsection{Ablation Experiments}
In the section, we will focus on the evaluation of the constraint effect of different objectives. To be specific, there are three types of loss functions: the contrast similarity loss, the domain adversarial training, and the time-invariant retrieval strategy, corresponding to different essential modules of ``CTVC''. Thus we retain the proposed model by discarding the term of some loss functions. We retrain our model without $\mathcal{L}_\text{adv-cls}$, and without Time-Invariant Retrieval~(TIR). Or, without $\mathcal{L}_\text{sim}$. 
As illustrated in Table~\ref{table:2}, when GRL layer is removed, the model failed in voice similarity score in VC task. It means the $\mathcal{L}_\text{adv-cls}$ objective matters in speaker identity eliminating. The proposed method also performs better than the model without TIR in speaker similarity, which indicates that time-invariant retrieval improves the ability to represent speakers. Also, it's shown that with content feature at frame-level without compression module, the quality of speech reduces evidently.
\begin{table}[htbp]
   \centering
   \vspace{-1em}
   \caption{Evaluation results of the ablation studies.} 
    \label{table:2}
    \begin{tabular}{l c c c}
     \hline
     Method & MCD$\downarrow$ & SIM Score$\uparrow$ & MOS$\uparrow$\\
     \hline
     CTVC & 8.02 $\pm$ 0.15 & 0.67 $\pm$ 0.38 & 3.66 $\pm$ 0.54\\
     w/o $\mathcal{L}_\text{adv-cls}$  & 8.07 $\pm$ 0.10 & 0.57 $\pm$ 0.31 & 3.51 $\pm$ 0.49\\
     w/o TIR & 8.05 $\pm$ 0.27 & 0.59 $\pm$ 0.41 & 3.49 $\pm$ 0.20\\
     w/o $\mathcal{L}_\text{sim}$ & 9.09 $\pm$ 0.19 & 0.60 $\pm$ 0.36 & 3.19 $\pm$ 0.13\\
     \hline 
    \end{tabular}
    \vspace{-2em}
\end{table}

\section{Conclusion}

In this paper, a novel method named ``CTVC" is proposed that can disentangle content and speaker-related representations for voice conversion. Specifically, contrastive learning is used to heighten the association between the frame-level content embedding and linguistic information at phoneme-level. To extract the time-invariant speaker information a time-invariant retrieval is proposed. Evaluation results demonstrate that the proposed method outperform than previous studies with better the intelligibility and similarity during voice conversion.
\section{Acknowledgement}
This paper is supported by the Key Research and Development Program of Guangdong Province under grant No.2021B0101400003. Corresponding authors are Xulong Zhang, Ning Cheng from Ping An Technology (Shenzhen) Co., Ltd (zhangxulong@ieee.org, chengning211@pingan.com.cn).
\clearpage
\bibliographystyle{IEEEbib}
\bibliography{AAVC}

\end{document}